\documentclass[twocolumn,preprintnumbers,showkeys,amsmath,amssymb]{revtex4}
\usepackage{graphicx}

\begin{document}
\title{Single qubit quantum secret sharing with improved security}

\author{Guang Ping He}
 \email{hegp@mail.sysu.edu.cn}
\affiliation{School of Physics \& Engineering and Advanced
Research Center, Sun Yat-sen University, Guangzhou 510275, China\\
and Center of Theoretical and Computational Physics, The University
of Hong Kong, Pokfulam Road, Hong Kong, China}

\author{Z. D. Wang}
 \email{zwang@hkucc.hku.hk}
\affiliation{Department of Physics and Center of Theoretical and
Computational Physics, The University of Hong Kong, Pokfulam Road,
Hong Kong, China}

\begin{abstract}
Analyzing carefully an experimentally feasible non-entangled single
qubit quantum secret sharing protocol and its modified version
[Phys. Rev. Lett. 95, 230505 (2005); \textit{ibid}. 98, 028902
(2007)], it is found that both versions are insecure against
coherent attacks though the original idea is so remarkable. To
overcome this fatal flaw, here we propose a protocol with a distinct
security checking strategy, which still involves single qubit
operations only, making it possible to achieve better security of
quantum secret sharing with current technology.
\end{abstract}


\maketitle

\newpage

\section{Introduction}

Suppose that Alice has some secret data, e.g., the password for a
locker or the access code for a computer program. She wants her
employees to share the data so that it can be regained if and only
if all employees collaborate, while any subset of the employees
cannot do so successfully. This is a typical secret sharing problem.
Secret sharing is also an element for building up many other
complicated cryptographic protocols. With the fascinating
development of quantum cryptography, using quantum methods to
achieve secure secret sharing has caught great interests both
theoretically and experimentally
\cite{qi46,qi103,qi89,qi90,qi275,qi404}. However, most of the
quantum secret sharing (QSS) protocols have to rely on entangled
quantum states. So far, the preparation of entangled states has
still been quite inconvenient in practice, and it has been even
harder to store them for more than a brief time, especially when
different parts of the entangled states are shared and kept
separated by different participants. For this reason, even though
some of these protocols can be demonstrated in laboratory, they are
likely still far from practical applications.

Recently, a novel QSS protocol was proposed and experimentally
demonstrated by Schmid \textit{et al}. \cite{Schmid}, which involves
non-entangled single qubits only. Thus the protocol is very
promising in practice. However, a subtle security loophole was found
later \cite{Comment}.
A modification on the original protocol was then proposed by Schmid \textit{%
et al}. \cite{Reply}. But here it will be shown that the modified
one is even less secure than the original one. Most important, we
propose a novel protocol with a distinct security checking strategy.
It closes the existing security loopholes successfully, while the
great feasibility of the original protocol is still maintained.
Therefore, the present protocol is of significance both
theoretically and practically.

In the next section, we will briefly describe the original protocol
proposed in Ref. \cite{Schmid} and the corresponding cheating
strategy proposed in
Ref. \cite{Comment}. In Sec. 3, the modified protocol proposed in Ref. \cite%
{Reply} will be outlined, and our cheating strategy to this version
will be proposed. Then we will give a simple solution to these two
cheating strategies, and pinpointing out the remaining loopholes in
Sec. 4. Our novel protocol will be proposed in Sec. 5, and the
security analysis will be elaborated in Sec. 6.

\section{The original protocol and the corresponding cheating strategy}

First, we would like to note that the QSS protocols involved in this
paper all concentrate on the sharing of a single classical bit. But
in fact, such protocols can be used for sharing any kind of
information. For example, if Alice wants the other participants to
share a classical $m$-bit string, they can simply run the protocol
$m$ times. Or if they want to share a quantum state, Alice can
prepare a quantum system in this state, then applies on it an
unitary transformation $U_{x}\in \{U_{1},U_{2},...,U_{m}\}$, and
gives the quantum system to any one of the participants. Meanwhile,
she shares the classical data $x$ among all other participants with
the protocol. Then the secret quantum state can be recovered when
all other participants collaborate, while any subset of the
participants cannot know which unitary transformation is $U_{x}$, so
they cannot unlock the secret even if they own the corresponding
quantum system.

Now let us recall briefly the original QSS protocol proposed by
Schmid \textit{et al}. \cite{Schmid}. The goal of the proposal is
that: after each of $N$ participants inputs the secret data, any
$N-1$ participants should be able to infer the secret input of the
remaining participant if and only if they collaborate. That is, the
protocol has the flexibility that any one of the participants can be
regarded as Alice, while the other $N-1$ participants can be
regarded as the employees who share the secret data of Alice. Let
$\left\vert 0\right\rangle $ and $\left\vert 1\right\rangle $\
denote the two orthonormal states of a qubit. Define $\left\vert \pm
x\right\rangle =(\left\vert 0\right\rangle \pm \left\vert 1\right\rangle )%
\sqrt{2}$\ and $\left\vert \pm y\right\rangle =(\left\vert
0\right\rangle \pm i\left\vert 1\right\rangle )/\sqrt{2}$.
\textit{The original protocol} is actually stated as follows.

\textit{(1) The first participant }$R_{1}$\textit{\ prepares a
single qubit in the state }$\left\vert +x\right\rangle $\textit{. }

\textit{(2) The qubit is passed through the }$N$\textit{\
participants sequently. Each participant }$R_{j}$\textit{\
(}$j=1,...,N$\textit{) chooses
the secret data }$\varphi _{j}\in \{0,\pi /2,\pi ,3\pi /2\}$\textit{\ (}$%
R_{N}$\textit{\ can simply choose between }$0$\textit{\ and }$\pi /2$\textit{%
), and acts on the qubit with the unitary phase operator  }%
\begin{equation}
\hat{U}_{j}(\varphi _{j})=\left\{
\begin{array}{c}
\left\vert 0\right\rangle \rightarrow \left\vert 0\right\rangle  \\
\left\vert 1\right\rangle \rightarrow e^{i\varphi _{j}}\left\vert
1\right\rangle
\end{array}%
\right. .  \label{operator}
\end{equation}

\textit{(3) The last participant }$R_{N}$\textit{\ measures the
qubit in the basis }$\left\vert \pm x\right\rangle $\textit{. This
completes one run of the qubit communication. }

\textit{(4) Each participant divides his action for every run into
two
classes: a class }$X$\textit{\ corresponding to }$\varphi _{j}\in \{0,\pi \}$%
\textit{\ and a class }$Y$\textit{\ corresponding to }$\varphi
_{j}\in \{\pi /2,3\pi /2\}$\textit{. They broadcast the class of
their action for each run in random order, but keep the particular
value of }$\varphi _{j}$\textit{\ secret. }

\textit{(5) With the announced classification, they determine which
runs are valid runs that satisfy the condition }$\left\vert \cos (\sum\nolimits_{j}^{N}\varphi _{j})\right\vert =1$\textit{%
\ and lead to a deterministic measurement result of
}$R_{N}$\textit{. }

\textit{(6) The security check: the participants choose a subset of
valid runs, and announced the value of }$\varphi _{j}$\textit{\ of
each participant in random order. They make comparison between these
values and the measurement result in step (3) to detect cheating. }

\textit{(7) The task of secret sharing is achieved with the
remaining valid runs. When any subset of }$N-1$\textit{\
participants wants to infer the choice of }$\varphi _{R}$\textit{\
of the remaining participant, they reveal among themselves their
values of }$\varphi _{j}$\textit{. In the case in which this subset
contains the last participant, he reveals the measurement result in
step (3). }

In this protocol, a secret classical bit being shared can be encoded
with the secret input $\varphi _{j}$\ of each participant\
($j=1,...,N-1$). For
example, $\varphi _{j}=0$\ and $\varphi _{j}=\pi /2$\ can represent the bit $%
0$, while $\varphi _{j}=\pi $\ and $\varphi _{j}=3\pi /2$\ can
represent the bit $1$. The secret of the last participant $R_{N}$ is
his measurement result in step (3).

A significant merit of this protocol lies in that only the local
manipulation of phases on a communicated single qubit is needed, and
thus it is very feasible for practical realization. Unfortunately,
although the protocol can indeed accomplish the task that any $N-1$
participants are able to infer the choice of $\varphi _{R}$ of the
remaining participant if they collaborate, it was shown in Ref.
\cite{Comment} that a subset of less than $N-1$ participants may
also reach this goal in certain cases, namely, the protocol does not
reach the security it was supposed to have. The cheating strategy in
Ref. \cite{Comment} is outlined below.

\textit{Cheating strategy A: }

\textit{Suppose that the }$k$\textit{th participant }$R_{k}$\textit{\ (}$%
k\in \{2,...,N-1\}$\textit{) is cheating. After he receives the
qubit }$\chi _{k-1}$\textit{\ from }$R_{k-1}$\textit{\ in step (2),
he does not apply the
operator }$\hat{U}_{k}(\varphi _{k})$\textit{\ on it to obtain }$\chi _{k}$%
\textit{\ and pass it to }$R_{k+1}$\textit{. Instead, he keeps the
qubit unmeasured. He also prepares an EPR pair }$\left\vert \psi
\phi
\right\rangle =(\left\vert 00\right\rangle +\left\vert 11\right\rangle )/%
\sqrt{2}$\textit{, and sends }$\phi $\textit{\ to }$R_{k+1}$\textit{\ as }$%
\chi _{k}$\textit{\ while keeping }$\psi $\textit{\ to himself. In
step (4) when he broadcasts the class of his action, there can be
two cases:}

\textit{(i) Some }$R_{j}$\textit{\ (}$j<k$\textit{) has not
broadcast the class of his action yet. Then }$R_{k}$\textit{\ will
take no advantage from the cheating. But he will not be caught
either, because he can\ infer the class of the action corresponding
to }$\varphi _{k}$\textit{\ to announce\ by performing a collective
measurement on }$\chi _{k-1}$\textit{\ and }$\psi $\textit{\ in the
basis }$\{(\left\vert 00\right\rangle \pm \left\vert 11\right\rangle
)/\sqrt{2},(\left\vert 01\right\rangle \pm i\left\vert
10\right\rangle )/\sqrt{2}\}$\textit{. This will make }$\phi
$\textit{\
collapse to }$\hat{U}(\varphi _{k})\left\vert \chi _{k-1}\right\rangle $%
\textit{, where the value of }$\varphi _{k}$\textit{\ belongs to the class }$%
X$\textit{\ (or }$Y$\textit{)\ action if his measurement result is }$%
(\left\vert 00\right\rangle \pm \left\vert 11\right\rangle )/\sqrt{2}$%
\textit{\ (or }$(\left\vert 01\right\rangle \pm i\left\vert
10\right\rangle
)/\sqrt{2}$\textit{). He can further infer the exact value of }$\varphi _{k}$%
\textit{\ to announce in the security check in step (6). This is
because\ all participants have announced their classes of actions
before the check,
thus he can count the number of the class }$Y$\textit{\ action among }$R_{j}$%
\textit{\ (}$j<k$\textit{). If the number is even (or odd), he knows
that
the state of }$\chi _{k-1}$\textit{\ was }$\left\vert \pm x\right\rangle $%
\textit{\ (or }$\left\vert \pm y\right\rangle $\textit{) before he
measured it. Then he can calculate }$\varphi _{k}$\textit{\ by his
above result of the collective measurement on }$\chi
_{k-1}$\textit{\ and }$\psi $\textit{.}

\textit{(ii) All }$R_{j}$\textit{\ (}$j<k$\textit{) have broadcast
the classes of their actions. Then }$R_{k}$\textit{\ knows
}$E_{k-1}\equiv \left\vert \cos (\sum\nolimits_{j}^{k-1}\varphi
_{j})\right\vert $\textit{. By measuring }$\chi _{k-1}$\textit{\ in
the basis }$\{(\left\vert
0\right\rangle \pm i^{1-E_{k-1}}\left\vert 1\right\rangle )/\sqrt{2}\}$%
\textit{, he will know the state of }$\chi _{k-1}$\textit{. He also
measures
}$\psi $\textit{\ in the basis }$\{\left\vert \pm x\right\rangle \}$\textit{%
\ or }$\{\left\vert \pm y\right\rangle \}$\textit{\ to collapse }$\phi $%
\textit{\ into either }$\left\vert \pm x\right\rangle $\textit{\ or }$%
\left\vert \mp y\right\rangle $\textit{, and compares the result of }$\phi $%
\textit{\ with }$\left\vert \chi _{k-1}\right\rangle $\textit{\ to infer }$%
\varphi _{k}$\textit{\ to broadcast. Again, his cheating will not be
detected. But in this case he knows the exact state of }$\chi _{k-1}$\textit{%
\ and }$\chi _{k}$\textit{. Then a subset of less than
}$N-1$\textit{\ participants\ with }$R_{k}$\textit{\ included will
be able to infer the choice of }$\varphi _{R}$\textit{\ of another
participant.}

\section{The modified protocol and the corresponding cheating strategy}

Responding to Cheating strategy A, Schmid \textit{et al}. proposed a
modified protocol \cite{Reply}, where they attempt to ensure that
the second case never occurs and thus would make the cheating
futile.

\textit{Modified protocol 1: }

\textit{All steps are the same as those of the original protocol,
except that in step (4), the participants\ always announce the
classification in
the order }$R_{N}\rightarrow R_{N-1}\rightarrow ...\rightarrow R_{1}$\textit{%
. }

This modified protocol indeed evades the second case of the cheating
strategy, however, it is still insecure with even poor security, as
shown below.

\textit{Cheating strategy B: }

\textit{Similar to cheating strategy A, after the dishonest }$R_{k}$\textit{%
\ (}$k\in \{2,...,N-1\}$\textit{) receives the qubit }$\chi
_{k-1}$\textit{\ from }$R_{k-1}$\textit{,\ he prepares an EPR pair
}$\left\vert \psi \phi
\right\rangle =(\left\vert 00\right\rangle +\left\vert 11\right\rangle )/%
\sqrt{2}$\textit{, and sends the second qubit }$\phi $\textit{\ to }$R_{k+1}$%
\textit{\ as }$\chi _{k}$\textit{\ while keeping }$\psi $\textit{\
to
himself. But the difference is that he needs not to keep }$\chi _{k-1}$%
\textit{\ unmeasured in this strategy. Instead, he can measures it
immediately in either one of the basis }$\left\vert \pm x\right\rangle $%
\textit{\ or }$\left\vert \pm y\right\rangle $\textit{\ at his will.
}

\textit{In step (4) when }$R_{k}$\textit{\ broadcasts the class of
his action, all }$R_{j}$\textit{\ (}$j>k$\textit{) have broadcast
the classes of
their actions since it was thus suggested in modified protocol 1. }$R_{k}$%
\textit{\ counts the number of the class }$Y$\textit{\ action among }$R_{j}$%
\textit{\ (}$j>k$\textit{). If the number is even (or odd), he
measures the
qubit }$\psi $\textit{\ in the basis }$\left\vert \pm x\right\rangle $%
\textit{\ (or }$\left\vert \pm y\right\rangle $\textit{). Thus he
can infer the state of }$\phi $\textit{\ (i.e., }$\chi
_{k}$\textit{) from the entangled form of }$\left\vert \psi \phi
\right\rangle $\textit{. By comparing the state with the result in
his previous measurement on }$\chi _{k-1}$\textit{, he can always
infer }$\varphi _{k}$\textit{\ and finish the
rest of the protocol successfully, just as if he had applied the operator }$%
\hat{U}_{k}(\varphi _{k})$\textit{\ on }$\chi _{k-1}$\textit{\ and
passed it to }$R_{k+1}$\textit{\ without cheating. But unlike the
honest protocol, in
this case he always knows the exact states of }$\chi _{k-1}$\textit{\ and }$%
\chi _{k}$\textit{\ for any valid run. Therefore among the first }$k$\textit{%
\ participants (i.e., all }$R_{j}$\textit{\ with }$j\leq
k$\textit{), any subset of }$N^{\prime }=k-1$\textit{\ participants
with }$R_{k}$\textit{\ included can infer the choice of }$\varphi
_{R}$\textit{\ of the remaining
participant. Also, among the last }$N-k+1$\textit{\ participants (i.e., all }%
$R_{j}$\textit{\ with }$j\geq k$\textit{), any subset of }$N^{\prime }=N-k$%
\textit{\ participants with }$R_{k}$\textit{\ included can infer the
choice of }$\varphi _{R}$\textit{\ of the remaining participant.
Especially, if the participant }$R_{2}$\textit{\
(}$R_{N-1}$\textit{) cheats with this
strategy, he alone can always know the choice of }$\varphi _{1}$\textit{\ (}$%
\varphi _{N}$\textit{) of the first (last) participant. }

This strategy cannot be detected as shown below. Consider the case
where the number of the class $Y$ action among $R_{j}$ ($j>k$) is
even. According to the strategy, $R_{k}$ measures the qubit $\psi $\
in the basis $\left\vert \pm x\right\rangle $. If he has measured
the qubit $\chi _{k-1}$\ in the basis $\left\vert \pm x\right\rangle
$ too, he announces that he has applied a class $X$ action. Then
there can be two cases. (i) $\left\vert \pm x\right\rangle $ is the
wrong basis for measuring $\chi _{k-1}$. It means that the number of
the class $Y$ action among the partners $R_{j}$ ($j<k$)
is odd. Thus the total number of the class $Y$ action among all partners $%
R_{j}$ ($j\in \{1,...,N\}$) will be odd, i.e., $\cos
(\sum\nolimits_{j}^{N}\varphi _{j})=0$\ so that the corresponding
run will not be recognized as a valid run in step (5) of the
protocol. That is, even if $R_{k}$ has not cheated, the run will not
lead to a deterministic measurement result of $R_{N}$. Therefore
such a run cannot reveal $R_{k}$'s cheating. (ii) $\left\vert \pm
x\right\rangle $ is the correct basis for measuring $\chi _{k-1}$,
i.e., the number of the class $Y$ action among the partners $R_{j}$
($j<k$) is even. The corresponding run is then a valid run that may
be chosen for the security check. But since the measurement results
of $R_{k}$ on both $\chi _{k-1}$ and $\phi $\ (i.e., $\chi _{k}$)
are both correct, the value of $\varphi _{k}$ he inferred can surely
pass any check
successfully. On the other hand, if $R_{k}$ has measured $\chi _{k-1}$\ in $%
\left\vert \pm y\right\rangle $, he announces that he has applied a
class $Y$ action. Then if $\left\vert \pm y\right\rangle $ is the
wrong basis for
measuring $\chi _{k-1}$, the run will not be a valid one either. Or if $%
\left\vert \pm y\right\rangle $ is the correct basis, $R_{k}$'s
measurement results and inferred value are all correct. In either
case, $R_{k}$'s cheating can also escape from being detected.
Similar results can also be
found in the case where the number of the class $Y$ action among $R_{j}$ ($%
j>k$) is odd.

Therefore, the modified protocol 1 is not secure either. Moreover,
unlike the case of the original protocol, where the cheating
strategy A can only gain information on the secret data in certain
cases, the cheating strategy B can always be successful for the
modified one. In addition, the cheating strategy B is more feasible
to be implemented, because no collective measurements on $\chi
_{k-1}$ and $\psi $ are required, while they are needed in the
cheating strategy A. The cheater here even needs not to store the
qubit $\chi _{k-1}$ for a long period of time. Consequently, the
modified protocol 1 proposed in Ref. \cite{Reply} is even less
secure than the original one.

\section{A simple solution and remaining loopholes}

If we want to defeat only the above cheating strategies alone, we
would have the following simple solution.

\textit{Modified protocol 2: }

\textit{All steps are the same as those of the original protocol,
except that in step (4), the order in which the participants
announce the
classification can be arbitrary, as long as }$R_{1}$\textit{\ and }$R_{N}$%
\textit{\ are always the last two to announce. }

With this modification, when $R_{k}$ ($k\in \{2,...,N-2\}$) is to
broadcast the class of his action in step (4), the cases where
either all $R_{j}$ ($j>k $) or all $R_{j}$ ($j<k$) have broadcast
the classes of their actions will never occur. Thus the protocol is
made secure against the above two cheating strategies.

Unfortunately, there are still some serious drawbacks in this one.
First, it
cannot stand the multi-cheater attack. Suppose that there are two cheaters, $%
R_{k}$ ($k\in \{2,...,N-2\}$) and $R_{1}$, and they exchange
information secretly. Then $R_{k}$ can still cheat with the strategy
A in the case where the participant $R_{j}$ ($2\leq j<k$) have
broadcast the classes of their actions before $R_{k}$ does, because
$R_{1}$ can tell $R_{k}$ his choice beforehand. For the same reason,
when $R_{k}$ and $R_{N}$ are cheaters, the cheating strategy B still
works in the case where $R_{j}$ ($k<j\leq N-1$) have broadcast the
classes of their actions before $R_{k}$ does. Similarly, even if we
put further restrictions on the order of the broadcast of the
classes of actions (e.g., by fixing the order as $R_{N/2}\rightarrow
R_{N/2+1}\rightarrow R_{N/2-1}\rightarrow R_{N/2+2}\rightarrow
...\rightarrow R_{1}\rightarrow R_{N}$ when $N$ is even), it is
still possible to cheat if more participants are dishonest. Second,
none of the above three versions is able to locate the cheater(s).
Even if the participants find disagreed announcement in the security
check, what they know is merely the existence of cheater(s), but
they never know exactly who is (are) cheating. Because of this, a
dishonest participant surely inclines to cheat as he is benefitted
from a successful cheating while has no risk to be caught, even if
the cheating fails.

\section{Our improved protocol}

To close these security loopholes completely, below we present a
protocol with a distinct security checking strategy.

\textit{Our improved protocol: }

\textit{(I) The }$N$\textit{\ participants agree on the number
}$n$\textit{\
of the total runs\ of the qubit communication, a weight }$w$\textit{\ (}$%
w>0.75n$\textit{\ is recommended), and a binary linear }$(n,l,d)$\textit{%
-code }$C$\textit{\ \cite{code}, where the minimal distance
}$d$\textit{\
between every pair of codewords satisfies }$%
n-w^{2}/n-l/4-lw/(2n)+3lw^{2}/(4n^{2})<d<2(n-w)$\textit{\
\cite{note}. Each
participant }$R_{j}$\textit{\ (}$j=2,...,N-1$\textit{) chooses secretly an }$%
n$\textit{-bit codeword }$c_{j}=(c_{j1}c_{j2}...c_{jn})$\textit{\ from }$C$%
\textit{\ whose weight (the number of }$1$\textit{\ in }$c_{j}$\textit{) is }%
$w$\textit{. }

\textit{(II) For }$i=1$\textit{\ to }$n$\textit{: }

\textit{\qquad (II-1) }$R_{1}$\textit{\ prepares a single qubit in
the state }$\left\vert +x\right\rangle $\textit{. }

\textit{\qquad (II-2) The qubit is passed through the }$N$\textit{\
participants sequently. Each participant }$R_{j}$\textit{\ (}$j=1,...,N$%
\textit{) applies an action on the qubit. For }$R_{j}$\textit{\ (}$%
j=2,...,N-1$\textit{), the action is chosen according to the
}$i$\textit{th
bit\ of the codeword }$c_{j}$\textit{. If }$c_{ji}=0$\textit{, }$R_{j}$%
\textit{\ applies a class }$X$\textit{\ or }$Y$\textit{\ action; else if }$%
c_{ji}=1$\textit{, }$R_{j}$\textit{\ applies a class }$Z$\textit{\ action. }$%
R_{1}$\textit{\ and }$R_{N}$\textit{\ apply the class }$X$\textit{\ or }$Y$%
\textit{\ action only. %Here, the class }$X$\textit{\ and }$Y$\textit{\
%actions are defined as the same as those in the original protocol. That is, }%
%$R_{j}$\textit{\ chooses the secret data }$\varphi _{j}\in \{0,\pi \}$\textit{%
%\ (class }$X$\textit{) or }$\varphi _{j}\in \{\pi /2,3\pi /2\}$\textit{\
%(class }$Y$\textit{) and acts on the qubit with the unitary phase operator }$%
%\hat{U}_{j}(\varphi _{j})$.
The class }$Z$\textit{\ action means that }$R_{j}$\textit{\ chooses
the secret data }$\varphi _{j1}\in \{0,\pi /2\}\ $\textit{and acts
on the qubit with }$\hat{U}_{j}(\varphi _{j1})$\textit{,\ then
measures the qubit in the basis }$\left\vert \pm x\right\rangle
$\textit{. He then chooses another secret data }$\varphi _{j2}\in
\{0,\pi /2,\pi ,3\pi /2\}\ $\textit{and acts on the measured qubit
with }$\hat{U}_{j}(\varphi _{j2})$\textit{,\ and sends the qubit to
the next participant. }

\textit{\qquad (II-3) }$R_{N}$\textit{\ measures the qubit in the basis }$%
\left\vert \pm x\right\rangle $\textit{. }

\textit{(III) For }$i=1$\textit{\ to }$n$\textit{, each }$R_{j}$\textit{\ (}$%
j=2,...,N-1$\textit{) announces the bit }$c_{ji}$\textit{\ in random
order. Thus at the end of this stage, all codewords
}$c_{j}$\textit{\ (}$j=2,...,N-1 $\textit{) are publicly aunnounced.
}

\textit{(IV) Security check 1: the participants check whether each }$c_{j}$%
\textit{\ (}$j=2,...,N-1$\textit{) is a codeword from }$C$\textit{\
with the weight }$w$\textit{. }

\textit{(V) For }$i=1$\textit{\ to }$n$\textit{, the participants
who announced }$c_{ji}=0$\textit{\ broadcast the classes of their
actions in random order. }

\textit{(VI) The participants determine which runs are valid runs,
i.e., the runs which contain no class }$Z$\textit{\ action\ and
satisfy }$\left\vert \cos (\sum\nolimits_{j}^{N}\varphi
_{j})\right\vert =1$\textit{. }

\textit{(VII) Security check 2: the participants choose a subset of
valid runs and all the runs containing the class }$Z$\textit{\
action. For each valid run, each participant announces his choice of
}$\varphi _{j}$\textit{\ in random order. They compare these values
with the measurement result in
step (II-3) to detect cheating. For each run containing the class }$Z$%
\textit{\ action, the participants who applied the class
}$X$\textit{\ or }$Y $\textit{\ action (except }$R_{1}$\textit{\ and
}$R_{N}$\textit{) announce
their choices of }$\varphi _{j}$\textit{\ in random order. Then }$R_{1}$%
\textit{, }$R_{N}$\textit{\ and these who applied the class
}$Z$\textit{\
action announce (in any order) all their choices of }$\varphi _{j1}$\textit{%
\ (including }$\varphi _{N}$\textit{) and the results of the
measurement
first, and then all }$\varphi _{j2}$\textit{\ (including }$\varphi _{1}$%
\textit{), and check whether they are in agreement with the
announcements of the other participants. That is, they use the
measurement result and the value of }$\varphi _{j2}$\textit{\
announced by a participant who applied the class }$Z$\textit{\
action and the values of }$\varphi _{j}$\textit{\ announced by these
who applied the class }$X$\textit{\ or }$Y$\textit{\ action to infer
what measurement result should be found by the next participant who
applied the class }$Z$\textit{\ action, and check whether this
participant indeed found this result. }

\textit{(VIII) When no disagreement is found, the task of secret
sharing is
thus achieved with any of the remaining valid runs. When any subset of }$N-1$%
\textit{\ participants want to infer the choice of }$\varphi
_{R}$\textit{\
of the remaining participant, they reveal among themselves their values of }$%
\varphi _{j}$\textit{\ and the measurement result in step (II-3) if
the last participant is included in this subset. }

\section{The security analysis}

Obviously this protocol can accomplish the task that any subset of
$N-1$ participants can infer the choice of $\varphi _{R}$ of the
remaining participant if they collaborate. Also, it will be proven
here that the protocol can meet the requirement that any subset of
less than $N-1$ participants cannot infer the choice of $\varphi
_{R}$ of a honest participant with the above cheating strategies.
The main reason is that the introduction of the class $Z$ action
provides a physical approach for honest participants to check the
communication channel between them, so that limits are put to the
cheaters' freedom on applying the cheating action. Meanwhile, the
codeword method adopted in the protocol can further give rigorous
mathematical guarantee that the remaining freedom of the cheater
will result in zero knowledge of the secret data.

Now we elaborate the proof in detail. Let us consider the most
severe case where $N-2$ participants are cheaters. Let $R_{a}$ and
$R_{b}$ ($a<b$) denote the rest two honest participants. Consider
first the case where $R_{a} $ and $R_{b}$ are neighbors in the qubit
transmission, i.e., $b=a+1$. Obviously the other $N-2$ cheaters can
infer the value of $\cos (\varphi _{a}+\varphi _{b})$ only. But if
they want to know each secret data $\varphi _{a}$ or $\varphi _{b}$\
alone, they must perform a man-in-the-middle eavesdropping attack
between $R_{a}$ and $R_{b}$. However, when the number $n $ of the
total runs of qubit communication is sufficiently large, there will
be plenty of runs in which both $R_{a}$ and $R_{b}$ applied the
class $Z$ action, and the bases in which $R_{a}$ sending the qubit
and $R_{b}$ measuring the qubit are exactly the same. When there is
no eavesdropper presented, the announcement of $R_{a}$ and $R_{b}$
in security check 2 will match with each other. Else if there is an
eavesdropper between them who
intercepts and measures the qubit sent by $R_{a}$, and then resends it to $%
R_{b}$, mismatched announcement will occur with a non-vanishing
probability in each of these runs. Then the probability for the
eavesdropper to escape the detection will drop exponentially to zero
as $n$ increases. Therefore, to infer any one of $\varphi _{a}$ or
$\varphi _{b}$, the $N-2$ cheaters still need to collaborate with
either $R_{b}$ or $R_{a}$, i.e., the collaboration of $N-1$
participants is needed so the protocol is secure.

Secondly, consider the remaining case where $R_{a}$ and $R_{b}$ are
separated. Since the goal of the cheaters is to learn the state of
the qubit sent by $R_{a}$ (or received by $R_{b}$) so that $\varphi
_{a}$ (or $\varphi _{b}$) can be inferred, at least one cheater
$R_{k}$ ($a<k<b$) between $R_{a} $ and $R_{b}$ must replace the
honest action with a cheating one (otherwise the participants are
acting honestly and there is no cheating at all). No matter what
strategy the cheaters may use, it seems that this cheating action
should have the following general features. To learn the state of
the qubit sent by $R_{a}$, $R_{k}$ must stop the qubit instead of
passing it to the next participant, so that he can perform
appropriate operations later; and to continue with the protocol,
$R_{k}$ must prepare and send the next participant another qubit,
which may entangled with other qubits or systems owned by himself or
other cheaters. The above cheating strategies A and B are both such
examples. Meanwhile, to pass security checks, the cheating action
must be able to be announced as a class $X$, $Y$ or $Z$\ action when
needed. But this cannot always be done, as proven by the following
four steps.

(a) \textit{To cheat successfully, the cheater }$R_{k}$\textit{\
should be
able to announce the cheating action as a class }$X$\textit{\ or }$Y$\textit{%
\ action.} An honest class $X$ or $Y$ action cannot be announced as a class $%
Z$\ action, because the participant cannot announce the result of
the measurement on the qubit he received correctly since he did not
keep it. Therefore, although a cheating action can always be
announced as a class $Z$\ action, if it cannot be announced as a
class $X$ or $Y$ action, $R_{k}$ must apply the honest class $X$ or
$Y$ action for exactly $n-w$ runs to pass security check 1. Then for
any valid run, the action of $R_{k}$ is honest, cheating being
impossible.

(b) $R_{k}$\textit{\ cannot announce the cheating action as a class }$X$%
\textit{\ or }$Y$\textit{\ action in a run where both }$R_{a}$\textit{\ and }%
$R_{b}$\textit{\ applied the class }$Z$\textit{\ action, or he will
be detected in security check 2 with a non-trivial probability.} To
announce the cheating action as a class $X$ or $Y$ action correctly,
$R_{k}$ needs to know $\varphi _{k}$ exactly. The value of $\varphi
_{k}$\ will be affected by the choice of $\varphi _{j}$\ of any
single participant $R_{j}$\ ($j\neq k $). Thus it can be determined
only when $R_{k}$\ knows the classes of actions of all $R_{j}$\
($a\leq j<k$) or all $R_{j}$\ ($k<j\leq b$). Therefore strategies A
and B seem to be the only two cheating strategies for $R_{k}$ to
infer $\varphi _{k}$. But if both $R_{a}$\ and $R_{b}$ applied the
class $Z$ action, they never need to announce their choice before
$R_{k}$ announces $\varphi _{k}$ in security check 2. Thus $R_{k}$
will not have enough information to calculate the correct value of
$\varphi _{k}$, so he has to announce it by guess and stands a
non-trivial probability to be detected.

(c) \textit{The number of runs in which }$R_{k}$\textit{\ can
announce the cheating action as a class }$X$\textit{\ or
}$Y$\textit{\ action without being detected is not sufficient for
}$R_{k}$\textit{\ to pass security check 1.}{\normalsize \ }

From point (b), it is seen that $R_{k}$ can safely announce the
cheating action as any of the class $X$, $Y$ or $Z$ action freely
only in the runs where he is sure that either $R_{a}$\ or $R_{b}$
applied a class $X$ or $Y$ action. Now let us evaluate the number of
these runs.

In a $(n,l,d)$-code $C$, the number of possible codewords having the weight $%
w$\ grows when $n$\ increases while $l/n$\ and $w/n$\ are fixed and
$d<2(n-w) $. Therefore when only a few\ bits of the codeword used by
$R_{a}$\ or $R_{b} $\ are disclosed, $R_{k}$\ cannot know the rest
of the codeword with certainty. Suppose that $R_{k}$\ can identify
the codeword used by $R_{a}$\ or $R_{b}$\ only after $l^{\prime }$
bits are revealed. Then in each of the
first $l^{\prime }$ runs, $R_{k}$ cannot know the choices of $c_{ji}$ of $%
R_{a}$\ and $R_{b}$ before they announce. Since in step (III)
$c_{ji}$ is announced in a random order, there are three
possibilities: both $R_{a}$\
and $R_{b}$ announced before $R_{k}$ does; one and only one of $R_{a}$\ and $%
R_{b}$ announced before $R_{k}$ does; none of $R_{a}$\ and $R_{b}$
announced before $R_{k}$ does. In the case where the cheater can get
the most benefit, the probabilities for these cases to occur are
$1/4$, $1/2$, and $1/4$, respectively (a rigorous calculation on
these probabilities is provided in the appendix). In the first two
cases $R_{k}$ can be sure of the class of action of $R_{a}$\ and/or
$R_{b}$. Since each participant should apply the class $Z$ action
$w$ times, the probability for $c_{ai}=0$\ (or $c_{bi}=0$) is
$1-w/n$. Therefore, the probability for such a run to be the one in
which $R_{k}$ can announce the cheating action as a class $X$ or $Y$
action is
\begin{equation}
p_{1}=(1-w^{2}/n^{2})/4+(1-w/n)/2.  \label{p1}
\end{equation}

In each of the last $n-l^{\prime }$ runs, suppose that the choice of
$c_{ji}$ ($i\in \{l^{\prime }+1,..,n\}$) of each participant can be
inferred by the others from the announced $c_{ji}$ ($i\in
\{1,..,l^{\prime }\}$) of the
first $l^{\prime }$ runs. Thus $R_{k}$ knows the choice of action of $R_{a}$%
\ and $R_{b}$ before they announce. The probability for such a run
to be the one in which $R_{k}$ can announce the cheating action as a
class $X$ or $Y$ action is then
\begin{equation}
p_{2}=1-w^{2}/n^{2}.
\end{equation}

Totally, even if $R_{k}$ applies the cheating action in all the $n$
runs, the maximum number of runs in which he can announce the
cheating action as a class $X$ or $Y$ action is
\begin{eqnarray}
n^{\prime } &=&l^{\prime }p_{1}+(n-l^{\prime })p_{2}  \nonumber \\
&=&n-w^{2}/n-l^{\prime }/4-l^{\prime }w/(2n)+3l^{\prime
}w^{2}/(4n^{2}). \label{n'}
\end{eqnarray}

Therefore, after $R_{k}$ finishes applying all the actions (either
honest or cheating ones) on the $n$ qubits in step (II), the major
part of the $n$-bit string $c_{k}$ which he can announce in step
(III) is already determined.
Only $n^{\prime }$ bits at the most can be altered between $0$\ and $1$ by $%
R_{k}$\ according to the announcement of $R_{a}$\ and $R_{b}$, while
the other bits have to be announced honestly or announced as $1$. In
general, a
codeword of a binary linear $(n,l,d)$-code $C$ can be identified if $l$ or $%
n-d$ bits are revealed, and $l\leq n-d$\ holds for any non-trivial
binary
linear code \cite{note}. Therefore we can take $l^{\prime }=l$ in Eq. (\ref%
{n'}). Since it is suggested in the protocol to choose $%
d>n-w^{2}/n-l/4-lw/(2n)+3lw^{2}/(4n^{2})$, we have $d>n^{\prime }$.
Note that the distance between any codewords of code $C$ is not less
than $d$, altering less than $d$ bits of a codeword will not result
in another codeword. Therefore, among all the possible strings which
can be obtained by
altering no more than $n^{\prime }$ bits of $c_{k}$, only those less than $%
1/\left(
\begin{array}{c}
n-n^{\prime }/2 \\
n^{\prime }/2%
\end{array}%
\right) $ portion are valid codewords from $C$. Note that, which
string is the one that can finally be announced by $R_{k}$ is
determined by the
position of the runs in which $R_{a}$\ or $R_{b}$ announced either $%
c_{ai}=0\ $or $c_{bi}=0$ before $R_{k}$ announces $c_{ki}$. Due to
the random order of the announcement, $R_{k}$ will generally not be
so lucky that the final string which he can announce happens to be a
valid codeword while has also the weight $w$ exactly, except with a
probability which can be made arbitrarily small by increasing $n$.
Thus $R_{k}$ cannot pass security check 1 by altering $n^{\prime }$
bits of $c_{k}$.

(d) \textit{Seen from the above points, the probability for }$R_{k}$\textit{%
\ to cheat successfully is arbitrarily small as }$n$\textit{\
increases.} Point (c) ensures that $R_{k}$ is unable to cheat
successfully by altering his announcement on the choice of action
only in the runs in which $R_{a}$\ or $R_{b}$ announced either
$c_{ai}=0\ $or $c_{bi}=0$ before $R_{k}$ announces $c_{ki}$. On the
other hand, a dishonest $R_{k}$ may take the risk to announce the
cheating action as a class $X$\ or $Y$\ action in the runs where
both $R_{a}$\ and $R_{b}$\ applied the class $Z$\ action, so that
the announced string $c_{k}$ can be a valid codeword with the weight
$w$. But it was proven in point (b) that in any single run,
announcing the cheating action as a class $X$\ or $Y$\ action stands
a non-trivial probability to be detected. Let $\varepsilon $ denotes
this probability. Since the minimum distance $d$ between codewords
increases with $n$, the number of runs in which both $R_{a}$\ and
$R_{b}$\ applied the class $Z$\ action while $R_{k}$ needs to
announce the cheating action as a class $X$\ or $Y$\ action to make
$c_{k}$ valid will also grow as $n$ increases. Thus the total
probability
for $R_{k}$ to pass the whole protocol without being detected will be $%
(1-\varepsilon )^{O(n)}$, which can be made arbitrarily small by increasing $%
n$.

\section{Summary and discussions}

Thus we showed that our QSS protocol is secure against the above
known cheating strategies. Moreover, when $R_{a}$ and $R_{b}$
detected cheating, they know that the cheater is located between $a$
and $b$. In the case where $b=a+2$ or more participants are honest
and $n $ is sufficiently large so that the values of $a$ and $b$ run
through many different combinations, the cheater can be more
precisely located.

On the other hand, however, there may be other attacks that do not
have the features we considered above, thus may not be covered by
our security analysis. For example, in point (b) of our proof, we
assumed that
strategies A and B are the only two cheating strategies for inferring $%
\varphi _{k}$ perfectly. Though it looks correct so far, we have to
admit that a rigorous proof is lacking. Imperfect cheating
strategies which may only infer the value correctly with a small
probability is not considered either. Also, Eq. (\ref {n'}) is
calculated under the assumptions that the codewords of the
participants have no correlation with each other and the randomness
in the order of announcements satisfies the distribution shown in
the appendix. Is it possible that there are wiser collaboration
strategies for the cheaters to break these assumptions? What if they
agree on a course of action before the procedure starts? Or if they
use quantum states other than those specified in the original
protocol? They may even collaborate at the quantum level. That is,
they store their codewords with entangled quantum states so that
there is quantum correlation between their values and the
corresponding actions. Though there is no obvious sign showing that
these can lead to a successful cheating immediately so far, the
potential possibilities are numerous. It seems unpractical to expect
a security proof that is completely generally. Therefore it is worth
further studying the exact boundary of the security of our protocol
in future works.

Nevertheless, our protocol can stand more cheating strategies than
previous ones. We also wish to emphasize that, implementation of our
protocol involves merely single qubit operations, no entanglement is
needed. Thus it can easily be realized with the technology reported in Ref. \cite%
{Schmid}. Therefore we can enjoy the improved security with exactly
the same experimental setup.

It is also worth noting that as pointed out in Ref. \cite{qi103},
the task of secret sharing can also be achieved by combining
classical secret sharing with quantum key distribution (QKD)
protocols \cite{qi365}. More rigorously, the participant (Alice) who
holds the secret data can split the secret classically into many
shares, and transfer each share to each of the other participants
via a private channel using QKD. But this requires a private quantum
communication channel between Alice and every single one of the
other participants. That is, when $N$ participants want to share a
secret, it should be known beforehand who acts as Alice, then she
needs to set up a private channel with each of the rest $N-1$
participants, so totally $N-1$ channels are needed. If they want to
make it possible that any one of them can be Alice at any time, they
will need a private channel between every pair of them, i. e.,
totally $(N-1)N/2$ channels are needed. On the
contrary, in our protocol the participants can set up the secret keys $%
\varphi _{j}$ first, before deciding who acts as Alice. At a later
stage, anyone of them can be Alice, and encodes her secret with the
keys $\varphi _{j}$. All these are achieved by the $N-1$ symmetric
channels between them. Therefore our protocol is more flexible and
efficient, especially when the number of participants is large.

We thank Bang-Hai Wang and Qin Li for helping. The work was
supported by the NSFC under grant Nos.10605041 and 10429401, the RGC
grant of Hong Kong, the NSF of Guangdong province under Grant
No.9151027501000043, and the Foundation of Zhongshan University
Advanced Research Center.

\appendix
\section{}

Here we elaborate why the probabilities mentioned in point (c) of
Sec. 6 are $1/4$, $1/2$, and $1/4$, respectively. That is, we
evaluate the probabilities for the following three cases in step
(III) of our protocol:

(A) both $R_{a}$\ and $R_{b}$ announced before $R_{k}$ does;

(B) one and only one of $R_{a}$\ and $R_{b}$ announced before
$R_{k}$ does;

(C) none of $R_{a}$\ and $R_{b}$ announced before $R_{k}$ does.

Generally, when the sequence for each $R_{j}$\ ($j=2,...,N-1$) to
announce the bit $c_{ji}$\ in step (III) is perfectly random (i.e.,
any possible sequence will occur with exactly the same probability,
and the order of the occurrence of these sequences is completely
unpredictable), the probabilities for these three cases to occur
should all be $1/3$\ according to statistic theory. However, two
questions are raised: who creates these perfectly random sequences?
and how to do it?

Until today, it is still widely believed that unconditionally secure
quantum coin toss and related multi-party secure computation
protocols are impossible. Therefore in a multi-party cryptographic
protocol among the distrustful parties, it is hard to find a source
which creates perfectly random sequences trusty to all parties. To
circumvent this problem, a practical substitution can be done as
follows. For some or all of the parties, each suggests a sequence
for announcing $c_{ji}$, then other parties vote one-by-one either
to accept or to reject this sequence (rejection should be limited to
a finite times otherwise there may never reach any agreement). This
procedure is repeated until finally there are $n$ sequences that are
accepted by all parties.

Now let us see what strategy each party would use on deciding which
sequences is acceptable so that the finally sequences look secure to
him. According to the security analysis in Sec. 6, we can see that a
cheater prefers to announce later than the others do. Therefore each
party tends to accept a sequence in which his position is in the
rear. But if among the
resultant sequences, the probability for a specific party to announce $%
c_{ji} $\ later than the others is significantly larger than $1/2$,
the others may think that this party is trying to cheat so they will
reject these sequences. As a result of the trade-off, it seems
``fair'' for a party $R_{j_{1}}$ to accept the sequences having the
following property: for any
other party $R_{j_{2}}$, the two cases -- $R_{j_{1}}$ announces before $%
R_{j_{2}}$ does, and $R_{j_{2}}$ announces before $R_{j_{1}}$ does
-- should occur with equal probabilities \textit{in every
situation}. That is, these probabilities should be $1/2$ when the
statistical estimation is made not only on the entirety of all
accepted sequences, but also within any subset of the sequences
where the relative positions of other parties with respect to
$R_{j_{1}}$ are fixed (please see the three-party example below for
illustration).

When this strategy is adopted, one party can effectively prevent
others from taking too much advantages from the sequences of the
announcement, while without making himself looks like a cheater.
Nevertheless, the final accepted sequences will be biased from
perfect randomness. Consider the order of announcement of $R_{a}$,
$R_{b}$ and $R_{k}$ among many parties. In $50\%$ sequences,
$R_{a}$\ announced before $R_{k}$ does. Since in these sequences,
the probabilities for $R_{b}$\ to announce before or after $R_{k}$
does should both be $1/2$, there will totally be $25\%$ sequences in
which both $R_{a}$\ and $R_{b}$ announced before $R_{k}$ does. In
the other $50\%$ sequences, $R_{a}$\ announces after $R_{k}$ did.
Since in these sequences, the probabilities for $R_{b}$\ to announce
before or after $R_{k}$ does should both be $1/2$, there will
totally be $25\%$ sequences in which none of $R_{a}$\ and $R_{b}$
announced before $R_{k}$ does. This is why in point (c) of Sec. 6,
the probabilities for cases (A) and (C) to occur are both $1/4$.
Thus the probability for case (B) to occur is $1-1/4-1/4=1/2$.

As an example, let us consider the case where there are only three parties $%
R_{a}$, $R_{b}$ and $R_{k}$. When the sequences is perfectly random,
the following six sequences (denoted by the indices of the parties)
should occur with equal probabilities: $abk$, $bak$, $akb$, $bka$,
$kab$, and $kba$. Then cases (A), (B) and (C) all occur with
probability $1/3$.

However, such a result of sequences does not satisfy the above
``fair'' condition. For instance, in $R_{a}$'s\ point of view,
whenever $R_{k}$\ announces after $R_{a}$ did (i.e., in $abk$,
$bak$, and $akb$), $R_{b}$\ has only probability $1/3$ (i.e., when
$bak$\ occurs)\ to announce before $R_{a}$ does, while having
probability $2/3$\ (i.e., when $abk$\ and $akb$\ occur) to announce
after $R_{a}$ did. Similarly, $R_{a}$\ and $R_{b}$\ will find the
sequences not being fair in every situation either. But if some of
the
sequences, e.g. $abk$ and $kba$,\ are rejected so that only $bak$, $kab$, $%
akb$, and $bka$ (note that these sequences have the feature that both $R_{a}$%
\ and $R_{k}$\ are presented in the middle twice) occur\ with equal
probability, while the probabilities for $abk$ and $kba$ to occur
are neglectable, then the result will be fair to both $R_{a}$\ and
$R_{k}$.\ Meanwhile in $R_{b}$'s point of view, though this result
does not satisfy the fair condition for him, it does not have to be
rejected since it still
looks random when estimated on the entirety, because the probabilities for $%
R_{b}$\ to be presented in the front or rear of these sequences are both $%
1/2 $. For the same reason, if only $abk$, $kba$, $akb$, and $bka$
(i.e., both $R_{b}$\ and $R_{k}$\ are presented in the middle of the
sequences twice) occur\ with equal probability while the other two
sequences seldom occur, the result will be fair to both $R_{b}$\ and
$R_{k}$ while still acceptable to $R_{a}$. Else if only $bak$,
$kab$, $abk$ and $kba$ occur\ with equal probability, the result
will be fair to both $R_{a}$\ and $R_{b}$ while still acceptable to
$R_{k}$. Therefore, as a result of the accept-reject trade-off, the
most probably accepted sequences that are fair to majority will have
the feature that the parties within this majority will be presented
mostly in the middle of the sequences, especially when the number of
parties is large and some of them may be dishonest. Note that even
though such a result is not fair enough to few of the parties, it is
still fine to them since the sequences still have some overall
randomness, and there does not seem to have any other better rule of
accept-reject to replace the above ``fair'' condition. Meanwhile,
such a result of sequences (e.g., $bak$, $kab$, $akb$, and $bka$) is
not only favored by a potential cheater (e.g. $R_{k}$), but also
favored by an honest party (e.g. $R_{a}$). Thus it cannot be used to
inferred which party is trying to cheat. Therefore without loss of
generality, we can assume that such a biased result of the sequences
is finally accepted, and the cheater is within the majority of the
parties which are benefitted. Consequently, the probabilities for
cases (A), (B) and (C) to occur are $1/4$, $1/2$ and $1/4$
respectively, as used in the proof in Sec. 6.

It is also worth pinpointing out that, even if a mechanism can be
found which can avoid the sequences from being biased by the above
strategy, or can even create perfectly random sequences for the
parties, our security analysis is still valid. This is because when
the sequences are perfectly random, the probabilities for cases (A),
(B) and (C) to occur are all $1/3$ (as it can be verified by
estimating over all the six possible sequences in the above
three-party example). Then $p_{1}$ in Eq. (\ref{p1}) will be
replaced by
\begin{equation}
p_{1}^{\prime }=(1-w^{2}/n^{2})/3+(1-w/n)/3.
\end{equation}
Since $p_{1}^{\prime }<p_{1}$, the maximum number of runs in which
the cheater $R_{k}$\ can announce the cheating action as a class $X$
or $Y$
action will be less than the number $n^{\prime }$\ obtained in Eq. (\ref{n'}%
). Thus the cheater is less benefitted from the perfectly random
order of announcement than the case we studied in Sec. 6. Therefore
our security analysis already covers this case and is more general.


\begin{thebibliography}{99}
\bibitem{note} The $(n,l,d)$-code $C$\ should also satisfy the general
bounds for any binary linear code \cite{code}, e.g., \ the Singleton
bound, Plotkin bound, and the Hamming bound, which require $l\leq
n-d+1$, $l\leq n-2d$, and $l\leq n-nH_{2}(d/2n)$ respectively when
$n\rightarrow \infty $, where $H(x)\equiv -x\log _{2}x-(1-x)\log
_{2}(1-x)$.\ It can be verified that when $w>0.75n$, valid values
exist for $l$, $d$ to satisfied these bounds and the requirement
suggested in step (I) of our protocol simultaneously.
\end{thebibliography}
\end{document}